\documentclass[conference]{IEEEtran}
\IEEEoverridecommandlockouts
\usepackage{cite}
\usepackage{amsmath,amssymb,amsfonts}
\usepackage{algorithmic}
\usepackage{graphicx}
\usepackage{textcomp}
\usepackage{xcolor}
\usepackage{paralist}
\usepackage{amsmath}
\usepackage[linesnumbered,ruled,vlined]{algorithm2e}

\SetCommentSty{mycommfont}
\def\BibTeX{{\rm B\kern-.05em{\sc i\kern-.025em b}\kern-.08em
    T\kern-.1667em\lower.7ex\hbox{E}\kern-.125emX}}

\newcommand\inv[1]{#1\raisebox{1.15ex}{$\scriptscriptstyle-\!1$}}

\begin{document}

\title{Improving Confidentiality for NFT Referenced Data Stores\\
}

\author{\IEEEauthorblockN{ Sarad Venugopalan and Heiko Aydt}
	\IEEEauthorblockA{\textit{Singapore-ETH Centre} \\
		sarad.venugopalan@sec.ethz.ch, aydt@arch.ethz.ch }
	
}


\maketitle

\begin{abstract}
A non-fungible token (NFT) references a data store location, typically, using a URL or another unique identifier. At the minimum, a NFT is expected to guarantee ownership and control over the tokenised asset. However, information stored on a third party data store may be copied and stolen. We propose a solution to give control back to the information owner by storing encrypted content on the data store and providing additional security against hacks and zero day exploits. The content on our data store is never decrypted or returned to its owner for decryption during rekeying. Also, the key size in our protocol does not increase with each rekeying. With this, we reduce the synchronisation steps and maintain a bounded key size. 
\end{abstract}


\section{Introduction}
\label{sec:confidentiality}
Protecting the information on NFT referenced data stores is a pertinent problem. This is because information on a third party data store is easily copied, and we are unable to protect it~\cite{Dipanjan2022}. It may result in theft, by issuing a fake NFT that points to a copy~\cite{Stephen2021}. 
There is a class of NFT applications that requires the information owner to retain both ownership and control (for its information stored on a data store).
 To achieve varying degrees of control, we make the distinction between  licensing and ownership sale. In the licensing business model, its consumers must pay the information owner by sending monies/crypto coins to the NFT smart contract, to allow retrieval of information (and a licence with its terms of use) from a data store. Paying a NFT supplies its consumer with a licence to use the information but not sell it. The ownership is retained by its owner. For example, any user may freely view a low resolution art image on a data store but is required to pay the NFT to view its licenced and watermarked high resolution image. This gives serious buyers enough information to decide whether to buy its watermark free high resolution image. The high resolution images (watermarked and non-watermarked) are stored encrypted on the data store. Ownership sales involves transferring of the underlying digital asset token on the blockchain and supplying both high resolution image decryption keys, to its new owner. This way the non-watermarked high resolution art image is never made public. An  escrow account may be set up on a smart contract to ensure this transaction is paid for and decryption keys received. 
 
 Another application is in the building and construction industry~\cite{Cetin2021}.
 It may be useful for city planners and analytics companies, to have knowledge of recyclable and reusable  material in a building~\cite{Honic2021}.
 This information may be collected and digitised by a building owner, and converted into a tokenised data asset~\cite{Hunhevicz2022}.
 A consumer of this information must pay the NFT to retrieve the requested information and licence from a data store.
 Other use cases may include tokenising the power (utility) bills of an apartment in a building. The monthly power bills may be retrieved from a data store by paying the corresponding NFT. This may be useful in giving insights such as --- ``do not rent west sun facing high floor apartments to reduce power consumption". 
 In another application, a NFT may point to the electric wiring diagram or the plumbing diagram for an apartment. For repairs, the corresponding NFTs may be paid to retrieve the required wiring diagrams. This would save the contractors time and effort attempting to deduce its location behind plastered walls. 
 Effectively, monetisation incentives the asset owner to digitise information and tokenise the asset, allowing valuable but hard to find information to be licensed or sold for  profit.
To licence or sell a tokenised asset, the information must not find its way into the public domain.
Also, we recognise plain-text information might be illegally sold by a past owner or a licensed consumer. We rely on the sales and licensing terms to discourage uncontrolled plain-text information dissemination.

An information owner may decide to host the data store herself on the internet, but this solution suffers from high costs. It is individually expensive to buy  server infrastructure and manage network downtime. Since, it may not be cost-efficient for the owner to be always-online,  she may decide to delegate this functionality to an online third party data store. However, the hosting data store  is able to view all the information on its storage. To resolve this, an owner may encrypt and deposit the information on the third party data store, and keep  the decryption keys separately on another online key store. Again, the key store might be compromised by an external adversary using a  zero day exploit and  steal its decryption keys.
To prevent this, we need to more than just encrypt the plain-text information (hereafter referred to as record). One of the options is to use a hardware security module (HSM)~\cite{Stieglitz2022}. A HSM may be efficiently used when the device is trusted. However, where trust is not fully explicit, a more benign solution is required. 
The rest of the paper is organised as follows. We describe the  solution outline in section~\ref{sec:proposedsolution}. The system architecture  is discussed in section~\ref{sec:archconf}. Section~\ref{sec:protocol} explains the confidentiality protocol and section~\ref{sec:discussion} discusses its security and speed optimisations.

\begin{figure*}
	\centering
	\includegraphics[width=0.75\linewidth]{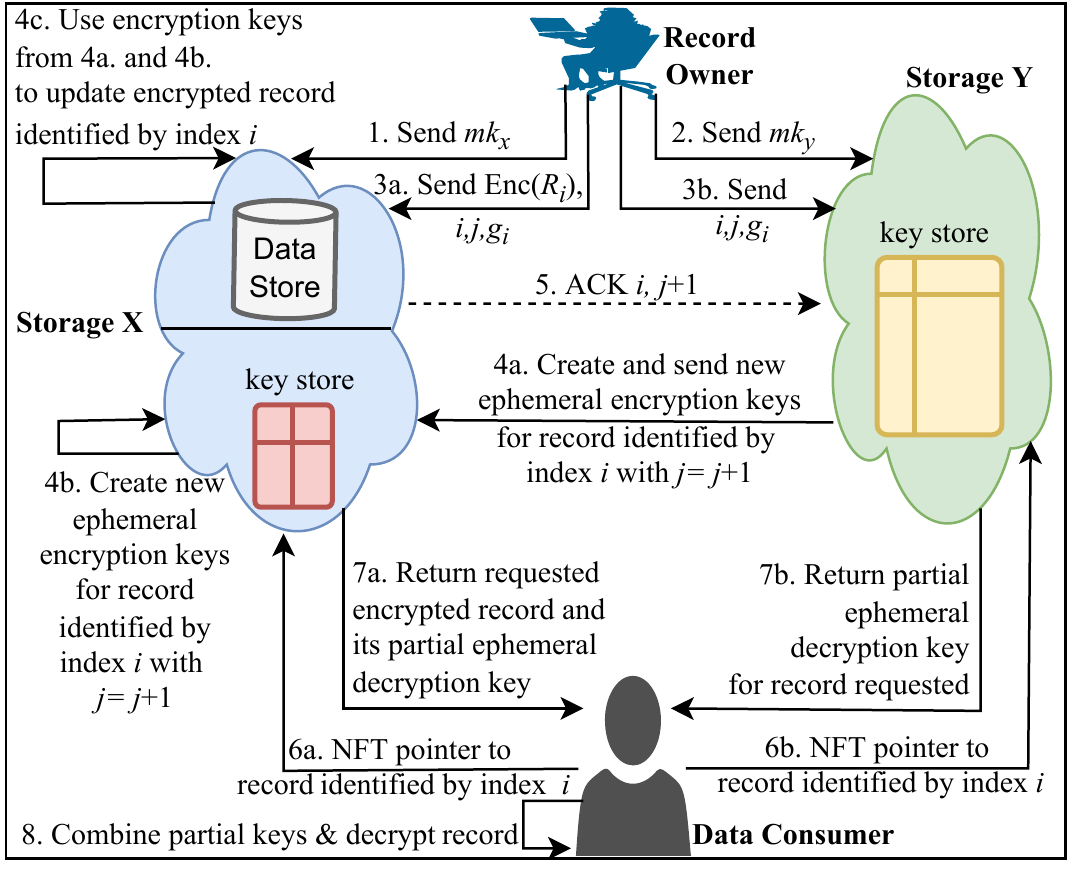}
	\caption{All communication channels between a sender and receiver are encrypted using public key cryptography. 
	The process involves bootstrapping the system (steps 1 \& 2), followed by adding the encrypted record to storage X's data store (step 3a.) and state information to Y (step 3b.). Re-encryption sequence for the encrypted record on storage X's data store is shown in step 4. An optional acknowledgement is seen in step 5. A consumer requesting and retrieving the encrypted record from the data store on storage X, and the keys required to decrypt it are shown in steps 6 to 8. 
	Note the partial decryption keys are provided to the consumer only after storage X and Y verifies the consumer paid the NFT smart contract for record use (not shown in figure). This is achieved by querying for payments to the NFT smart contract. All transactions on a blockchain are digitally signed and straightforward to verify.  }
	\label{fig:arch}
\end{figure*}

\section{Solution Outline}
\label{sec:proposedsolution}
We present a solution that involves no trusted third parties and uses ephemeral keys to encrypt and decrypt the records. Ephemeral keys improve protection against  zero day exploits that allow an external attacker to break in and steal keys. With ephemeral keys, any previously used (stolen) keys can not be used to decrypt the latest encrypted record on our data store. Our solution does not require the record owner to be always-online to supply decryption keys. The records on the data store are never decrypted and key updates do not require decryption. Our solution employs 2 key stores (see Fig.~\ref{fig:arch}), one is under the control of storage X (it also hosts the data store) and the other is  storage Y, rented by the information owner. We assume that  X and owner rented storage Y are collusion free.  X and Y are third party services that the owner is able to access via supported API calls. For example, X and Y may be online cloud hosting services.  X is unable to generate decryption keys for the records stored by itself. There are two secret master keys (MKs), one held by X and the other with Y. The content keys (CKs) are ephemeral keys used to encrypt the records. They change each time a record is served to a consumer. Each of X and Y hold partial content keys on their key store. They must be combined in order to encrypt/decrypt a record. Our solution partitions the storage of partial  keys and are held secret. We assume all communication channels are encrypted by default. I.e., each communicating party has access to the public key of the counterparty and uses public key cryptography. Hence, any information  sent to a consumer will be encrypted with her public key. Only the intended recipient is able to recover the plain-text record.
\section{System Architecture}
\label{sec:archconf}

\subsection{Stakeholders \& Threat Model}
\label{ssec:threatmodel}
The stakeholders are record owner, data consumer, third party storage  X  and Y. 
A record owner has ownership of the information. A data consumer requests this information to gain insights (or carry out analytics) by paying the required NFT smart contract.
X hosts a NFT referenced data store and a key store, whereas  Y only hosts a key store.
Both X and Y are assumed to be mutually non-trusting. For example, they are different hosting companies.
An adversary may eavesdrop on information passing through the communication channels.
Both X and Y are expected to carry out operations honestly but X may leak any plain-text data on its storage.
X or Y (but not both) may be compromised by an external adversary. 
\subsection{Components \& Interactions}
\label{ssec:compi}
Both X \& Y are access controlled. Only authorised users are able to view \& modify information on the data store and key stores.
A record owner  encrypts her plain-text records offline.
The bootstrap process is as follows (see Fig.~\ref{fig:arch}).
The record owner, after setting up system parameters 1.) sends a secret master key ($mk_x$) to  storage X's key store.
Further, the record owner 2.) sends another  secret master key ($mk_y$) to storage Y's key store.
This completes  the bootstrap process and storage X's data store is ready to receive encrypted records.
3a.) Record owner sends to storage X, $Enc(R_i), i, j, g_i$.
I.e., an encryption of plain-text  record $R$, uniquely identified by NFT index $i$. The value of $j$ corresponds to the number of times a record $R_i$ was encrypted. For its first encryption, the value of $j$ is 1.
The value of $g_i$ corresponds to the initial value of a pseudo random number generator (PRNG).
3b.) Record owner sends storage Y, $i, j, g_i$.
Next, the encrypted record on the data store  is updated (re-encrypted) as follows.    
4a.) Storage Y sends X, a new partial ephemeral encryption key (content key)  identified by index $i$. It also sends the updated counter $j+1$. 4b.) Storage X creates a new partial ephemeral encryption key for the updated $j+1$ counter. 4c.) Further, X uses the partial encryption key sent by storage Y along with its own newly generated partial encryption key, to update the encrypted record on its data store. 5.) An optional step is to acknowledge the updated counter $j+1$ for the record identified by $i$, to sync with storage Y.
Next, a consumer pays the required NFT smart contract for a record (not shown in Fig.~\ref{fig:arch}).
Further, the consumer
6a.) requests for an encrypted record identified by NFT index \textit{i}  from the NFT data store on X and 6b.) sends to storage Y, the NFT identifier of the record requested. 
Both X \& Y queries the NFT smart contract  to verify if the necessary payments were made for the record requested (not shown in Fig.~\ref{fig:arch}). Next, 7a.) Storage X returns the encrypted record and a partial ephemeral decryption key held by it. 
7b.) Storage Y returns its partial ephemeral decryption key. 8.) Consumer combines the partial ephemeral decryption keys to recover plain-text record $R_i$.
To ready  the next  consumer request for this record, step 4 of Fig.~\ref{fig:arch} is called to re-encrypt the record on the data store with a new pair of ephemeral keys.

\section{Confidentiality Protocol}
\label{sec:protocol}

Phases 1-3 are for bootstrapping the protocol and encrypting a record  offline (by its record owner).  Phases 4-6 corresponds to their online interactions. Phase 7 is the offline decryption of the record by its consumer. Phase 8 updates the record on the data store.
\subsubsection*{\textbf{Phase 1 (Setup Parameters)}}

A secret master key called $mk_x$ is generated and shared by the record owner directly with storage X (see step 1, Fig.~\ref{fig:arch}). It is a shared secret  known only to the owner and storage X. Another secret $mk_y$ is generated by the owner and shared with storage Y (see step 2, Fig.~\ref{fig:arch}). It is known only to her and storage Y.
The parameters for bootstrapping the confidentiality protocol are as follows:

Let $R=\{R_1,R_2,\ldots, R_n\}$ be the set of plain-text records.
The first step is to set up  a different generator for each $R_i\in R| {i\in \{1,\ldots, n\}}$ such that $g_i\in \mathbb{F}_p^*$, a prime field.
This is to initialise a PRNG with a large period.
Map each of the record onto  an element in $\mathbb{F}_p^*$ using an invertible map.
The value of $p$ is chosen to be a safe prime, i.e., $p=2\cdot  q +1$, where $q$ is a prime. Once chosen, $p$ remains unchanged throughout the protocol.
A safe prime is chosen to ensure the multiplicative group of order  $p - 1 = 2\cdot q$,  has no small subgroups that are non-trivial to detect.
Due to Fermat's little theorem~\cite{Koblitz}, to test  if any $a\in \mathbb{F}_p^*$ is a  generator, it is sufficient to verify if $a^{(p-1)/2} \equiv -1 \mod p$.
\footnote{Modular exponentiation by repeated squaring is used to compute $g^x$ $mod$ $p$. It has a time complexity of $\mathcal{O}$(($log$ $x$)$\cdot (log^2$ $p$))~\cite{Koblitz}. The increase in time complexity w.r.t. the exponent $x$ is logarithmic.}
Alternatively, to make sure that $a\in \mathbb{F}_p^*$ generates a large subgroup, it is sufficient to ensure $a^{2} \neq 1 \mod p$. Since, our data store may have millions of records, and a suitable $a_i$ (of large order/period) is required for each $R_i$, this is useful to quickly find an $a\in \mathbb{F}_p^*$, such that order$(a)=q$ or $2.q$.
Each elimination (of small subgroups) by testing requires only a single squaring operation modulo $p$.
When $q$ is chosen to be a sufficiently large prime, our generators $g_i$ may be substituted with $a_i$, since each of these elements generate a subgroup at least half the size of $p-1$.

\subsubsection*{\textbf{Phase 2 (Generate ephemeral encryption keys)}}
The record owner is required to encrypt her records before it is added to X's data store.
Let $||$ be the concatenation operator and $g_i$ be the generator corresponding to record $R_i$.
Owner carries out the following two sets of key generations for each  $R_i$ using $j=1$. 
For the record $R_i$, the owner computes  $ck^x_{i,j}$ and $ck^y_{i,j}$.
	\begin{multline}
		\label{eqn:eq2}
		ck^x_{i,j} = HMAC(mk_x, g_i^{1+2\cdot (j-1)}\mod p) ||\\ HMAC(mk_x, g_i^{2+ 2\cdot (j-1)}\mod p)
	\end{multline}
	
	\begin{multline}
		\label{eqn:eq3}
		ck^y_{i,j} = HMAC(mk_y, g_i^{1+2\cdot (j-1)}\mod p)||\\
		HMAC(mk_y,g_i^{2+ 2\cdot (j-1)}\mod p)        
	\end{multline}
	We use HMAC-SHA3-512 for hashing. It  generates a 512 bits output.
	Each of $ck^x_{i,j}$ and $ck^y_{i,j}$ (content keys) are a concatenation of 2 HMAC(.) outputs.
	Hence, $ck^x_{i,j}$ and $ck^y_{i,j}$ are each, typically, $1024$ bits long.
	The length of $mk_x$ and $mk_y$ (master keys),  are  each chosen to be 512 bits in length. I.e., same as the length of the HMAC output.
	We assume the safe prime $p$ chosen is  of length 1024 bits. The pair of ephemeral encryption keys for record $R_i$ are $ck^x_{i,j}\mod p$ and $ck^y_{i,j}\mod p$.
	
\subsubsection*{\textbf{Phase 3 (Encrypt a Record)}}
Record owner (on her offline computer) encrypts a plain-text record $R_i$. The offline record encryption uses $j=1$, for its first encryption. The arithmetic operations are in $\mathbb{F}_p^*$.

\begin{multline}
	\label{eqn:eq4}
	S_{i,1} = E(R_i) = ck^x_{i,1}\cdot ck^y_{i,1}\cdot R_i  
\end{multline}

\subsubsection*{\textbf{Phase 4 (Add an encrypted record to data store)}}
For the plain-text record $R_i$, record owner sends to storage X the values of $S_{i,1}, i, j=1, g_i$ (see step 3a, Fig.~\ref{fig:arch}).
The owner rented storage Y is sent the values of $i, j=1, g_i$ (see step 3b, Fig.~\ref{fig:arch}).

\subsubsection*{\textbf{Phase 5 (Re-encrypt record on data store)}}
Owner rented storage Y computes a new partial ephemeral key $ck^y_{i,j}$ by running Equation.~\ref{eqn:eq3} with $j\leftarrow j+1$ and sends it to storage X (see step 4a, Fig.~\ref{fig:arch} ).
Similarly, storage X computes $ck^x_{i,j}$ by running Equation.~\ref{eqn:eq2} with $j\leftarrow j+1$ (see step 4b, Fig.~\ref{fig:arch}).
Next, the data store on X re-encrypts its record $S_{i,j}$ (see Equation.~\ref{eqn:eq5}).
All arithmetic operations are in $\mathbb{F}_p^*$. This is  step 4c, in Fig.~\ref{fig:arch}.
\begin{multline}
	\label{eqn:eq5}
	S_{i,j} = {ck^x_{i,j}}\cdot {ck^y_{i,j}}\cdot S_{i,j-1} = {\prod_{k=1}^{j} ck^x_{i,k}\cdot  \prod_{k=1}^{j} ck^y_{i,k}}\cdot R_i
\end{multline}

\subsubsection*{\textbf{Phase 6 (Supply consumer with encrypted record and ephemeral decryption keys)}}

Consumer requests encrypted record and partial decryption keys (see step 6a and 6b, Fig.~\ref{fig:arch}).
Storage X looks up its data store to retrieve the latest $S_{i,j}$.
The partial ephemeral encryption key ${\prod_{k=1}^j ck^x_{i,k}}$
is constructed using Equation.~\ref{eqn:eq2}.
The partial ephemeral decryption key is trivially determined as
its multiplicative inverse, namely, $\inv{({\prod_{k=1}^j ck^x_{i,k}})} \mod p$. The latest encrypted record $S_{i,j}$ in the data store and its partial ephemeral decryption key is sent to the consumer (see step 7a, Fig.~\ref{fig:arch} ).
Storage Y carries out a similar set of operations to construct its ephemeral decryption key $\inv{({\prod_{k=1}^j ck^y_{i,k}})} \mod p$ for record $S_{i,j}$, using Equation~\ref{eqn:eq3}.
This ephemeral decryption key is sent to the consumer (see step 7b, Fig.~\ref{fig:arch}).

\subsubsection*{\textbf{Phase 7 (Record decryption by the consumer)}}
The consumer carries out the following computation to retrieve the plain-text record $R_i$ using its partial decryption keys (see Equation~\ref{eqn:eq6}).
Arithmetic operations are in $\mathbb{F}_p^*$.
This is step 8, in Fig.~\ref{fig:arch}.
\begin{multline}
	\label{eqn:eq6}
	R_i = D(S_{i,j}) = \inv{{\prod_{k=1}^{j} ck^x_{i,k}}} \cdot \inv{{\prod_{k=1}^{j} ck^y_{i,k}}} \cdot S_{i,j}
\end{multline}

\subsubsection*{\textbf{Phase 8 (Update the record on the data store)}}

The encrypted record $S_{i,j}$ on storage X is updated by re-encryption with a new pair of ephemeral keys. This is carried out by repeating phase 5 with an incremental value of $j$, corresponding to the record.

Consider the example shown in Fig.~\ref{fig:linear}.
Each record is first encrypted by its owner before it is added to  X's data store.
Plain-text records are $R_1, R_2$ and $R_3$ and the initial encrypted records on the data store are $S_{1,1}, S_{2,1}$ and $S_{3,1}$, respectively.
Here, $j=1$ corresponds to the initial encryption for the record.
On the data store, each encrypted record with $j=1$ is re-encrypted. This gives us $S_{1,2}, S_{2,2}$ and $S_{3,2}$.
In our example, an encrypted record for $R_3$ is requested by a consumer. The record served from the data store is $S_{3,2}$. 
The decryption of $S_{3,2}$ is $D(S_{3,2})=  ({ck^x_{3,1}}\cdot {ck^x_{3,2}})\cdot \inv{({ck^x_{3,1}}\cdot {ck^x_{3,2}})} \cdot ({ck^y_{3,1}}\cdot {ck^y_{3,2}}) \cdot \inv{({ck^y_{3,1}}\cdot {ck^y_{3,2}})} \cdot R_3 =  R_3$.
Once record $S_{3,2}$ is served to the consumer, the data store re-encrypts $S_{3,2}$ to give $S_{3,3}$.
\begin{figure}[h]
	\scriptsize
	\centering        
	\vspace{-0.2cm}
	\begin{equation}
		\boxed{
			\begin{array}{rcl}
				S_{1,1} = E(R_1)=  ({ck^x_{1,1}}\cdot {ck^y_{1,1}})\cdot R_1\\
				
				S_{2,1} = E(R_2)= ({ck^x_{2,1}}\cdot {ck^y_{2,1}})\cdot R_2\\
				
				S_{3,1} = E(R_3)= ({ck^x_{3,1}}\cdot {ck^y_{3,1}})\cdot R_3
				
				\\*
				\hline \\
				S_{1,2} =  \underline{ck^x_{1,2}}\cdot \underline{{ck^y_{1,2}}} \cdot ({ck^x_{1,1}}\cdot {{ck^y_{1,1}}})\cdot R_1\\
				
				S_{2,2} =  \underline{ck^x_{2,2}}\cdot \underline{{ck^y_{2,2}}} \cdot ({ck^y_{2,1}}\cdot {{ck^y_{2,1}}})\cdot R_2\\
				
				\triangleright S_{3,2} =  \underline{ck^x_{3,2}}\cdot \underline{{ck^y_{3,2}}} \cdot ({ck^x_{3,1}}\cdot {{ck^y_{3,1}}})\cdot R_3\\
				\hline\\
				S_{3,3} =  \underline{ck^x_{3,3}}\cdot \underline{ck^y_{3,3}}\cdot ({{ck^x_{3,2}}} \cdot {ck^y_{3,2}}\cdot {ck^x_{3,1}}\cdot {{ck^y_{3,1}}})\cdot R_3
			\end{array}
		}
	\end{equation}
	\vspace{-0.1cm}
	
	\caption{Encrypted records on a data store. All operations are modulo $p$. A record is served to a consumer only after its first re-encryption.    }
	\label{fig:linear}
\end{figure}

\section{Discussion and Practical Considerations}
\label{sec:discussion}
Generator $g_i$ is used as a PRNG to increase the hamming distance between subsequent variable inputs to the HMAC (as opposed to an incremental counter). The powers of the generator are the variable  input to the HMAC (see Equation.~\ref{eqn:eq2} and \ref{eqn:eq3}). The output of the HMAC is used as a cryptographically secure PRNG.
The security of the protocol relies on the difficulty to recover the secret master keys, $mk_x$ and $mk_y$, from its corresponding HMAC outputs.
Further, the security of HMAC used depends on the underlying hash algorithm, output size, and the key size~\cite{Bellare1996}.
Since we employ HMAC-SHA3-512 to compute $ck^x_{i,j}$ and $ck^y_{i,j}$, an adversary retrieving $mk_y$ from $ck^y_{i,j}$ is expected to be at least as hard as launching a first preimage attack on the SHA3-512 hash.
SHA3 uses Keccak~\cite{Bertoni2013} as its underlying algorithm and has so far shown excellent preimage attack resistance~\cite{KeccakHe2022,KeccakWang2022}.
Another possible attack is for storage X to attempt and infer the first ephemeral key $ck^y_{i,1}$ used in the  encryption of record $S_{i,1}$ (see Equation.~\ref{eqn:eq4}).
However, $ck^y_{i,1}$ is never sent to storage X by Y as part of the protocol (see Fig.~\ref{fig:arch}).
Storage Y sends the consumer, the inverse of its partial product of ephemeral keys, $(\prod ck^y_{i,k})\inv$ for the decryption of the record. At this point, the storage X and the consumer may collude to deduce the first ephemeral  key ($ck^y_{i,1}$) but this serves no useful purpose. Since all keys required for record decryption were received, the consumer may as well supply the plain-text record to storage X.
We do not attempt to prevent the dissemination of record information by the consumer, once it is decrypted. We rely on the data licence terms for the record usage to discourage the consumer from uncontrolled sharing of information.
With respect to computational speed, it is not necessary to regenerate past ephemeral keys and multiply them every time a record  decryption is required.
Computing the  product (in Phase 6) requires iterating over all values of $j$.
This may be sped up by storing the partial products modulo $p$, on the  storage.
Further, all keys and their products are computed modulo $p$. Hence, the encryption and decryption keys are bounded by the size of prime $p$.
\section{Conclusions}
\label{sec:conclusions}
We presented a protocol to improve the confidentiality of information stored on a third party data store. 
By using  two key stores, one alongside the data store on storage X and the other on owner controlled storage Y --- a high level of information confidentiality was achieved. The ephemeral keys used made it less vulnerable to hacks. It  may serve as a valuable tool for business owners to  control and selectively disseminate their content stored  on a third party data store.

\section*{Acknowledgment}

This research is supported by the National Research Foundation, under its Campus for Research Excellence and Technological Enterprise (CREATE) Programme.

\bibliographystyle{IEEEtran}
\bibliography{ref-reduced}

\end{document}